\documentclass[final,5p,times,twocolumn,sort&compress]{elsarticle}
\usepackage[dvipsnames]{xcolor}
\usepackage{graphicx}
\usepackage{changepage} 
\usepackage{amssymb}
\usepackage{amsmath}
\usepackage{ulem}
\usepackage{caption} 
\usepackage{multirow} 
\usepackage{hyperref}
\usepackage{float}
\usepackage{caption}
\usepackage{colortbl}

\usepackage[capitalise,noabbrev,nameinlink]{cleveref}
\hypersetup{
citecolor=blue,
 colorlinks=true,
 linkcolor=blue,
 filecolor=magenta,
 urlcolor=cyan,
}

\Crefname{equation}{Eq.}{Eqs.}

\begin{document}
\begin{frontmatter}

\title{Effects of hyperon potentials and symmetry energy in quark deconfinement}

\author{Rajesh Kumar}
\ead{rkumar6@kent.edu}
\author{Krishna Aryal}
\author{Alexander Clevinger}
\author{Veronica Dexheimer}
\ead{vdexheim@kent.edu}

\address{Department of Physics, Kent State University, Kent, Ohio, 44242, USA}

\begin{abstract}
In this letter we discuss how the results of recent nuclear experiments that correspond to measurements at low densities can affect the equation of state at large densities and temperatures, changing the particle composition and ultimately influencing  deconfinement to quark matter. In particular, saturation values of the hyperon potentials affect the hyperon content, while the symmetry energy at saturation directly regulates how the stiffness of the equation of state changes with isospin. We make use of a chiral model that describes nucleons, hyperons, and quarks to show how astrophysical conditions, such as the ones in neutron stars, present the ideal ground to study the effects of these two quantities in dense matter. In this case, for small charge fraction/ large isospin asymmetry, the couplings that reproduce different symmetry energy slopes can significantly modify deconfinement, with quantitative changes in the critical chemical potential depending on the deconfining potential. On the other hand, different values of the parameter that controls the hyperon potentials (kept within a range close  to experimental data) do not affect deconfinement significantly. 
\end{abstract}


\end{frontmatter}

\section{Introduction}
Recent nuclear experiments have reopened the discussion of the effects of strangeness and isospin in dense matter. First, the updated Pb Radius Experiment (PREX-II) reported on the neutron skin thickness $R_{\rm{skin}}$ \cite{PREX:2021umo}. This is directly related to the increase in energy for isospin-asymmetric systems, such as neutron stars. The difference between the energy per nucleon of fully asymmetric and symmetric matter at saturation density $n_{\rm{sat}}$, is referred to as the symmetry energy $E_{\rm{sym}}$, or $S$. But, most importantly, its derivative with respect to (baryon number) density, $L$, determines how the symmetry energy increases as the system becomes more dense with baryons or quarks, e.g., going towards the center of a neutron star. There is a strong correlation between $R_{\rm{skin}}$ and $L$ \cite{Furnstahl:2001un,Chen:2010qx} that allows neutron skin thickness to be used to infer values of symmetry energy. One issue is that, while PREX-II points to high values $L=(106\pm37)$ MeV \cite{Reed:2021nqk}, another recent PREX-II study of parity-violating asymmetry provides a much smaller value of $L=(54\pm8)$ MeV \cite{Reinhard:2021utv}. Furthermore, neutron skin thickness calculated from the proton elastic scattering (RHIC) and ultra-relativistic nuclear collisions (LHC) of $^{208}$Pb  can also provide a constraint on $L$, resulting in $L=60 \pm30$ MeV~\cite{Zenihiro:2010zz} and $L=79 \pm39$ MeV~\cite{Giacalone:2023cet}, respectively. In fig.~5 of Ref.~\cite{Li:2019xxz} the authors provide a compilation of terrestrial and astrophysical analyses of $L$. Results for the neutron skin thickness also relate to the tidal deformability $\tilde\Lambda$ of neutron stars, and larger values of $R_{\rm{skin}}$ correspond to $642\lessapprox\tilde{\Lambda}\lessapprox955$ \cite{Reed:2021nqk}, in tension with values estimated from LIGO (with minimal assumptions) $\tilde{\Lambda}\le630$ \cite{LIGOScientific:2018hze}.

Different experiments taking place at CERN have recently brought back the discussion of hyperon potentials. In relativistic models, the hyperon potentials are defined as $U_H=$ vector interaction $+$ scalar interaction, usually discussed at $n_{\rm{sat}}$ for isospin-symmetric matter. Since the emulsion and bubble chamber experiments from the 1980's, it has been known that the $\Lambda$-nucleon potential is negative with $U_\Lambda\sim-28$ MeV \cite{Millener:1988hp}, with more recent estimates obtaining slightly lower values $-32<U_\Lambda<-30$ MeV \cite{Fortin:2017dsj}. A couple of decades later, potentials for the $\Sigma$ and $\Xi$ hyperons were also measured. It was found that the $\Sigma$ potential is repulsive, estimated using measurements at KEK (the High Energy Accelerator Research Organization in Japan) to be $U_\Sigma=30\pm20$ MeV \cite{Gal:2016boi}, and that the $\Xi$ potential is attractive, measured by KEK and J-PARC (the Japan Proton Accelerator Research Complex) to be $U_\Xi=-21.9\pm0.7$ MeV. The latter is much lower than what had been previously found, e.g., $U_\Xi=-14$ MeV \cite{AGSE885:1999erv,Gal:2016boi}. Finally, a much less attractive potential was found from p-$\Xi$ correlation functions from the ALICE collaboration using three-momenta measured at center of mass energy $s=\sqrt{13}$ TeV \cite{Fabbietti:2020bfg,ALICE:2020mfd}. No error was provided, but their potential $U_\Xi=-4$ MeV agrees with the one calculated by the HAL-QCD collaboration. The latter used hyperon interactions extracted from (2+1)D lattice QCD in a Brueckner-Hartree-Fock (BHF) calculation to obtain $U_\Xi=-4$ MeV, in addition to $U_\Lambda=-28$ MeV and $U_\Sigma=+15$ MeV, with a statistical error of approximately $\pm2$ MeV \cite{Inoue:2019jme}.

The effects of hyperon potentials \cite{Iida:1998pi,Carroll:2008sv,Zschiesche:1999gf,Papazoglou:1997uw,Wang:2001hw,Weissenborn:2011kb,Petschauer:2015nea,Wang:2002pza,Kumari:2020mci,Fortin:2020qin,Schaffner-Bielich:2000igu,Hu:2021ket,Fu:2022eeb,Kochankovski:2023trc,Gusakov:2014ota,Chatterjee:2015pua,Isaka:2017nuc,Zhao:2017nlw,Djapo:2008au} and symmetry-energy slope \cite{Miyatsu:2022wuy,Lopes:2023dnx,Drischler:2013iza,Horowitz:2014bja,Trautmann:2017xlx,Sotani:2015lya,Providencia:2013dsa,Jiang:2012zs,Steiner:2004fi,Millerson:2019jkg,Negreiros:2018cho,Li:2020ass,Most:2021ktk,Yang:2023ogo,Hutauruk:2023mjj,Krastev:2021reh,Thapa:2021syu,Ghosh:2022lam,Suzuki:2022mow,Char:2023fue,Tsang:2023vhh,Lim:2019som,Chen:2015gba} beyond saturation have already been explored substantially. In this work we go beyond prior analyses, by considering their effects on quark deconfinement at finite temperature and out of weak ($\beta$) equilibrium. We make use of the Chiral Mean Field (CMF) model, which includes a consistent treatment of deconfinement to build multidimensional QCD (or high-energy) phase diagrams. We investigate parametrizations and couplings that reproduce different hyperon potentials and symmetry-energy slopes at saturation and determine what they imply for quark deconfinement. We also explore different deconfinement potentials that have different dependence on the baryon chemical potential  to account for the unknown strength of the deconfinement phase transition at low and intermediate temperatures.  We do this for different temperatures, charge fractions, and strangeness constraints. 
The structure of our paper is as follows.  In \cref{sec:formalism}, we briefly summarize the formalism adopted to describe dense matter. In \cref{sec:RnD}, we describe and discuss our results. Finally, we present our conclusions in \cref{sec:conclusion}.   

\section{Formalism}
\label{sec:formalism}

To study dense matter, we employ the Chiral Mean Field (CMF) model. It is a relativistic SU(3) effective chiral model, which has been modified to describe quark degrees of freedom, in addition to baryons. As a consequence, it allows for different strengths of the deconfinement phase transition to be reproduced, including a smooth crossover at large temperatures \cite{Dexheimer:2009hi}. 
The Lagrangian density is given by 
\begin{equation}
\mathcal{L} = \mathcal{L}_{\rm{Kin}} + \mathcal{L}_{\rm{Int}} + \mathcal{L}_{\rm{Self}} + \mathcal{L}_{\rm{SB}} - U_{\Phi} ,
\end{equation}
where the first term is the hadron and quark kinetic-energy density. The mesons do not contribute kinetically due to the use of the mean-field approximation. The other terms include an interaction term between the baryons or quarks and the (mean-field) mesons, a self-interaction term for the scalar and vector mesons, an explicit symmetry-breaking term, and a potential for the deconfining order parameter. The baryons included are nucleons and  hyperons, the quarks are the $3$ light ones, and the mesons are: the scalar-isoscalars $\sigma$ and $\zeta$, scalar-isovector $\delta$, vector-isoscalars $\omega$ and $\phi$, and vector-isovector $\rho$. Both $\zeta$ and $\phi$ are strange quark-antiquark states.

In particular, adding a separate adjustable mixed vector-isovector $\omega\rho$ interaction in the self-interaction term of vector mesons (first introduced in a Walecka-type model~\cite{Horowitz:2002mb}) can directly influence the symmetry-energy slope, as demonstrated for zero temperature $\beta$-equilibrated hadronic matter for the CMF and other relativistic models \cite{Dexheimer:2018dhb}. It has the form
\begin{equation}
\mathcal{L}_{\omega\rho} = g_{ \omega\rho} \omega_\mu \omega^\mu \rho_\mu \rho^\mu,
\end{equation}
where $g_{\omega\rho}$ is the coupling term for $\omega\rho$ interaction. The stronger the coupling, the softer the isospin-asymmetric equation of state (EoS, pressure vs. energy density) at low/intermediate densities, and the lower the symmetry-energy slope. This interaction also lowers the radii of intermediate-mass ($\sim1.4$ M$_\odot$) neutron stars and their tidal deformabilities, while not affecting their masses substantially \cite{Dexheimer:2018dhb}. One can also add higher-order vector interactions of the form
\begin{equation}
 \mathcal{L}_{\omega^4} =  g_{\omega^4}(\omega_\mu \omega^\mu)^2,
\end{equation}
where $g_{\omega^4}$ is a coupling constant with negative value. This correction can compensate for the small decrease in neutron-star mass (resulting from the previous interaction) by stiffening the EOS at large densities and, as a consequence, increasing the mass of heavy neutron stars. This has been shown for the CMF model for zero-temperature $\beta$-equilibrated matter \cite{Dexheimer:2020rlp}  and has the opposite behavior of what was shown e.g., in Ref.~\cite{Fattoyev:2010rx}, where the coupling constant $g_{\omega^4}$ was taken to be positive.

In the CMF formalism, the degrees of freedom change from hadronic to quark as the baryon chemical potential $\mu_B$ or temperature $T$ increase. This is because the effective masses of baryons $B$ and quarks $q$, in addition to being generated by the scalar mesons (or quark condensates), also depend on the deconfinement order parameter $\Phi$
\begin{align}
M_{B}^* &= g_{B \sigma} \sigma + g_{B \delta} \tau_3 \delta + g_{B \zeta} \zeta + M_{0_B} + g_{B \Phi} \Phi^2,  \nonumber \\
M_{q}^* &= g_{q \sigma} \sigma + g_{q \delta} \tau_3 \delta + g_{q \zeta} \zeta + M_{0_q} + g_{q \Phi}(1 - \Phi),
\end{align}
and on a mass correction $M_{0_i}$, which contributes to the masses of nucleons, hyperons, and quarks as follows
\begin{align}
M_{0_N} &=m_0 = 150~\rm{MeV}, \ \quad M_{0_H}= m_0- m_3 \left( \sqrt{2} \sigma_0 + \zeta_0  \right), \nonumber \\ 
M_{0_u}&= M_{0_d}=5~\rm{MeV}, \quad M_{0_s}=150~\rm{MeV},
\end{align}
with $m_0$ being a bare mass term for baryons, $M_{0_q}$ being the Higgs contribution to the quarks masses, and $m_3$ being an explicit breaking symmetry term (hereon referred to as ExpSB). Furthermore, similar to the Polyakov loop dynamics \cite{Fukushima:2003fw}, as $\Phi$ changes $0\to 1$, the effective masses of baryons become very large and the quarks become low, signaling deconfinement. 

\begin{figure}[t!]
\centering
\includegraphics[scale=0.35]{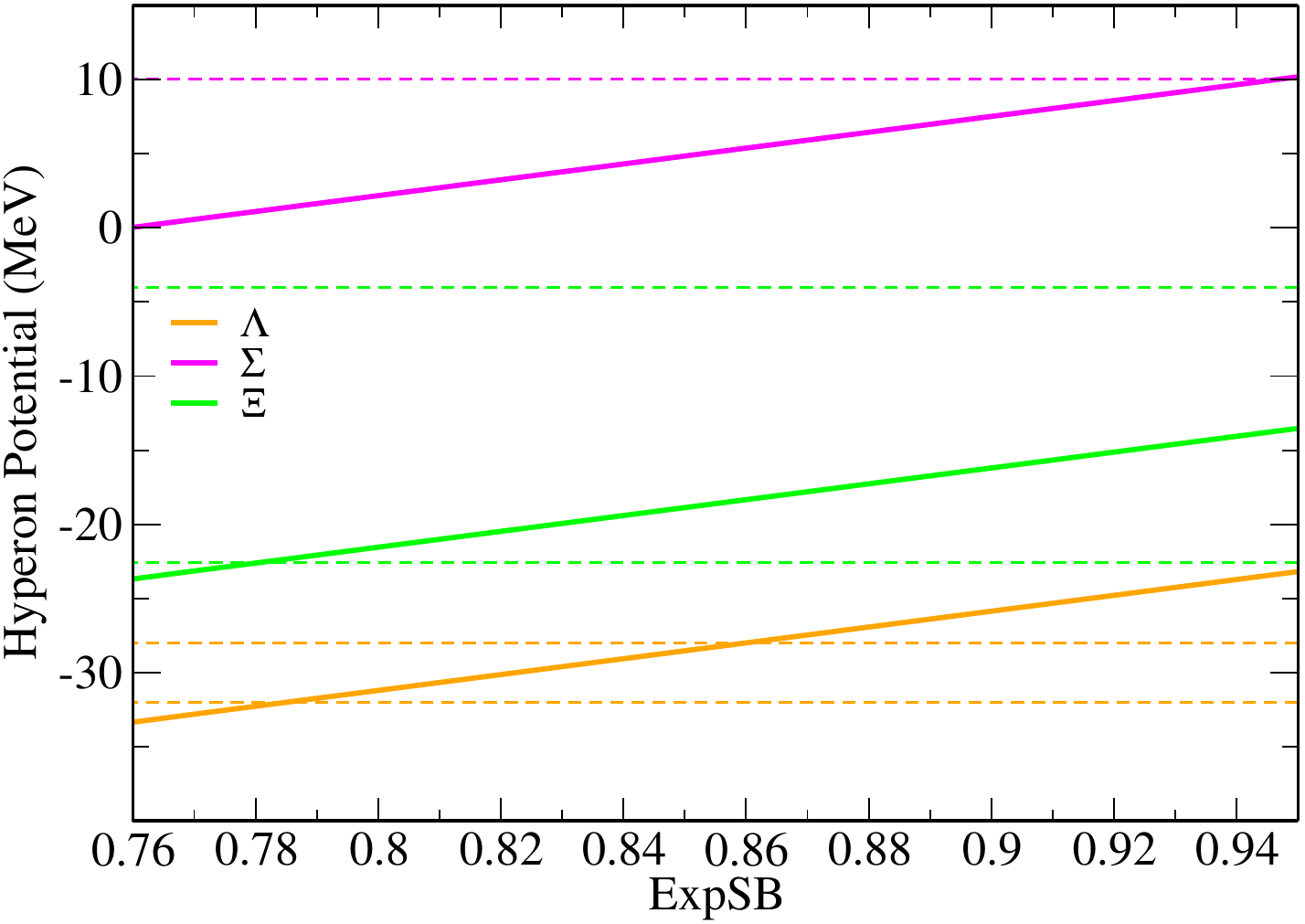}
\caption{Hyperon potentials as a function of explicit symmetry-breaking coupling for isospin-symmetric matter, $Y_Q=0.5$, at zero temperature at saturation. Horizontal lines show constraint ranges discussed in the introduction.}
\label{fig:HP_vs_esb3}
\end{figure}

The original CMF potential for $\Phi$ is
\begin{align}
U_{\Phi}&=\left(a_0T^4+a_1\mu_B^4+a_2T^2\mu_B^2\right)\Phi^2\nonumber \\ &+a_3T_0^4\ln\left(1-6\Phi^2+8\Phi^3-3\Phi^4\right),
\label{eq:U_Phi}
\end{align}
where $a$'s and $T_0$ are parameters fitted to constraints from the QCD phase diagram \cite{Dexheimer:2009hi}. We also make use of a new potential 
\begin{align}
U'_{\Phi}&=\left(a'_0 T^4+a'_1\mu_B^2+a'_2 T^2\mu_B^2\right)\Phi^2\nonumber \\ 
&+a_3 T_0^4\ln\left(1-6\Phi^2+8\Phi^3-3\Phi^4\right),
\label{eq:U_Phi'}
\end{align}
used with a different $g'_{B\Phi}$, $g'_{q\Phi}$. $U'_{\Phi}$ that was derived in Ref.~\cite{Dexheimer:2020rlp} and later used in Ref.~\cite{Clevinger:2022xzl}, in both cases only at zero temperature for $\beta$-equilibrated matter. It was shown that the weaker dependence on density or $\mu_B$ gives rise to a weaker first-order phase transition and the vector interactions for the quarks allow for pure quark matter to be present in stable hybrid stars without resorting to mixtures of phases.

\section{Results and Discussions}
\label{sec:RnD}

We start by illustrating for the first time the influence of the dimensionless parameter ExpSB on the hyperon potentials for isospin-symmetric matter at $n_{\rm{sat}}$.  Usually, it is fixed to ExpSB~$=0.86$ in order to produce $U_\Lambda=-28$ MeV, but in this work we vary ExpSB from $0.76$ to $0.95$ in order to vary $U_\Lambda$ by $\pm  5$ MeV. \cref{fig:HP_vs_esb3} shows how all three hyperon potentials change linearly with the change in ExpSB, with $U_\Lambda$ and $U_\Xi$  remaining negative and $U_\Sigma$ remaining positive.  The values in \cref{fig:HP_vs_esb3} lie within or are in the vicinity of predictions from theory and experiments (indicated by horizontal lines in the figure). Note that the only experimental value for $U_\Sigma$ has a very large error bar, of $40$ MeV, because it is based on a compilation of experimental and theoretical results. Its range has the same sign but it is more repulsive than our results.

\begin{figure}[t!]
\centering
    \includegraphics[scale=0.35]{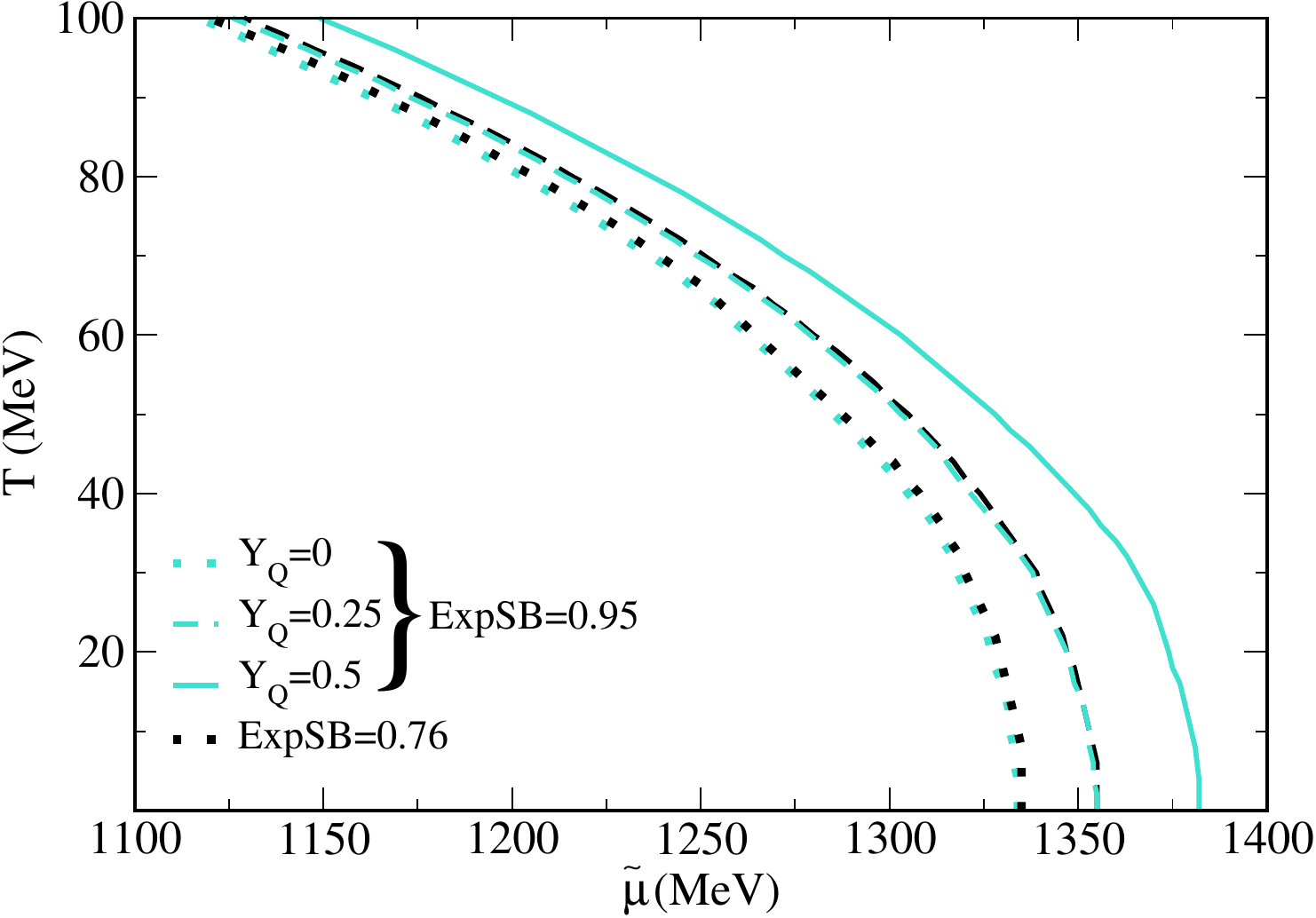}
  \caption{Temperature vs free energy phase diagram for matter without constraints on strangeness shown for different charge fractions and explicit symmetry-breaking couplings. The full lines overlap.}
  \label{fig:Tvsmub_yq0-yq050}
\end{figure}

But, more importantly than studying specific values, we are interested in determining if changes in ExpSB (that imply changes in hyperon potentials) affect particle composition, specifically quark deconfinement.  In addition, we would like to, for the first time, study ExpSB effects at finite temperature.  To do so, we built QCD phase diagrams for different conditions (corresponding to laboratory vs. astrophysics), easily identified by the degree of isospin asymmetry or electric charge and strangeness. In particular, we work with unitless fractions, such as the charge fraction (per baryon)
\begin{equation}
Y_Q = \frac{Q}{B} = \frac{\sum_i Q_i n_i}{\sum_i Q_{B_i} n_i} ,
\label{6h}
\end{equation}
where $Q_i$ is the electric charge, $n_i$ the density, and $Q_{B_i}$ the baryon number of each particle (1 for baryons and 1/3 for quarks). In our figures we plot the results as a function of free energy (per baryon) defined  for the cases we study as 
\begin{equation}
\tilde{\mu} = \mu_B + Y_Q\mu_Q +  Y_S\mu_S,
\label{eqn:muhat1}
\end{equation}
where $\mu_Q$ is the charge chemical potential (equivalent to the isospin chemical potential in our formalism, defined as the difference between the proton and neutron chemical potentials - see discussion in Section II B and Appendix A of Ref.~\cite{Aryal:2020ocm}) and $\mu_S$ the strange chemical potential. $Y_S$ is the strangeness fraction (per baryon)
\begin{equation}
Y_S = \frac{S}{B} = \frac{\sum_i S_i n_i}{\sum_i Q_{B_i} n_i} ,
\end{equation}
where $S_i$ is the strangeness of each particle. In the particular case of highly isospin-asymmetric matter without any constraint on strangeness, $Y_Q=0$ and $Y_S\neq 0$ (similar to astrophysical conditions, with $\mu_S=0$), or isospin-symmetric matter with zero net strangeness, $Y_Q=0.5$ and $Y_S=0$ (similar to laboratory conditions, with $\mu_Q=0$), one has $\tilde{\mu}=\mu_B$. 


\begin{figure}[t!]
\centering
  \includegraphics[scale=0.35]{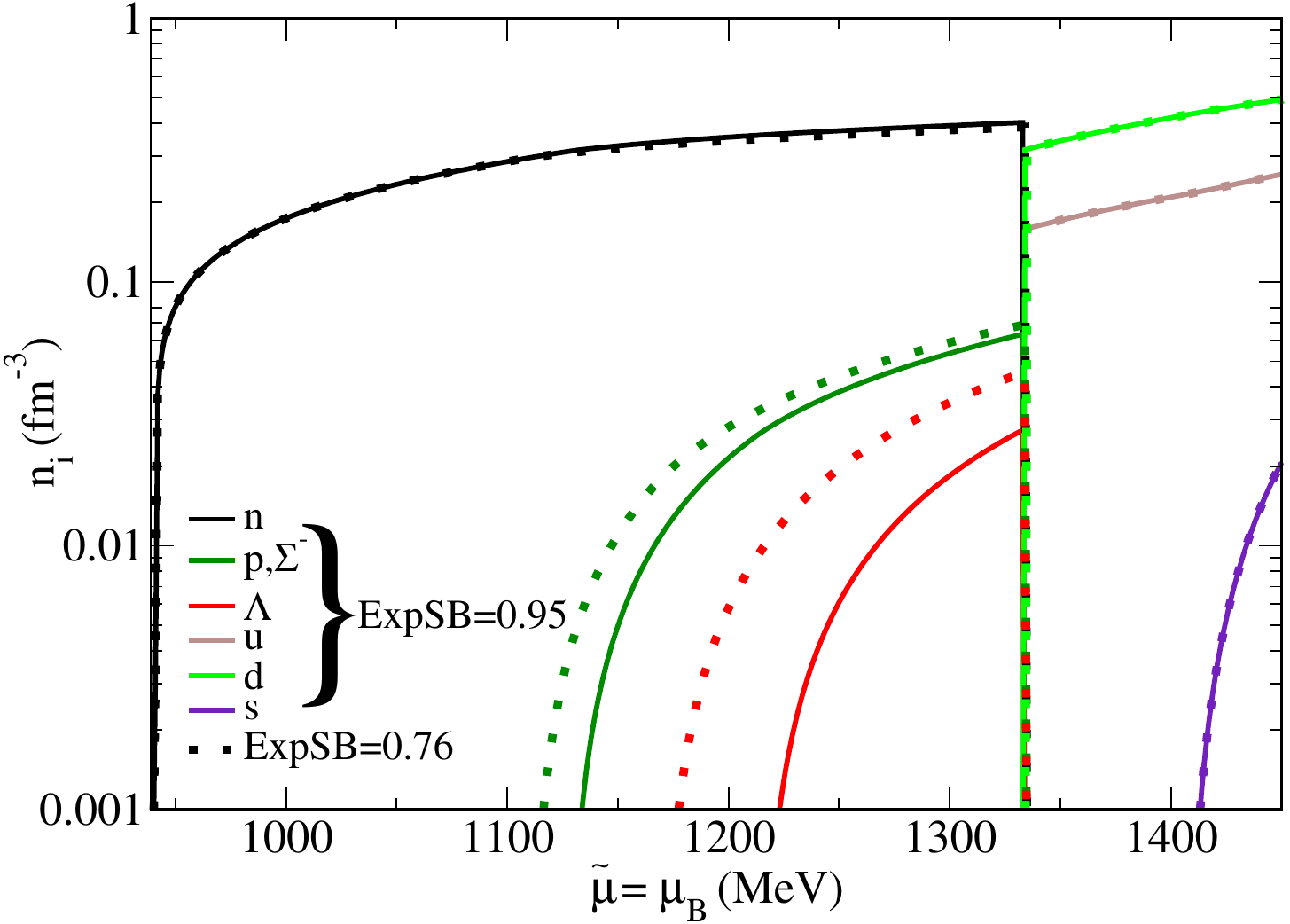}
  \caption{Particle population as a function of free energy (or baryon chemical potential) for zero-temperature, charge fraction $Y_Q=0$ matter with no constraints on strangeness shown for different values of explicit symmetry-breaking couplings. Quark densities were divided by $3$.}
  \label{fig:population}
\end{figure}

Starting with $Y_S\neq 0$, which is relevant for astrophysical time scales, in \cref{fig:Tvsmub_yq0-yq050} we present the changes in the deconfinement phase coexistence line that arise when ExpSB is varied within the range shown in \cref{fig:HP_vs_esb3}. The difference is very small, only perceptible for the cases of low $Y_Q$ (dotted lines), when the hyperons are favored, and does not depend significantly on $T$. 
%
To understand this result, we need to look at the particle content reproduced by these parameters. \cref{fig:population} depicts the particle population density for the case of $Y_Q=0$ and $T=0$. In the hadronic phase only neutrons, protons, $\Lambda$'s, and $\Sigma^-$'s appear, being the other hyperons suppressed by the first-order phase transition to quark matter, with up- and down-quarks, and later the strange one. Protons and $\Sigma^-$'s overlap due to imposing charge neutrality. A larger value of ExpSB enhances the values of the hyperon potentials, making the $\Lambda$'s and the $\Sigma$'s more massive, more energetically costly to produce, and, as a consequence, suppressed. Matter with fewer hyperons is stiffer and, consequently, the phase transition takes place at lower energy densities (both on the hadronic and quark sides), which correspond to lower baryon densities (both sides), and slightly lower chemical potentials or free energies. 

But note that $Y_Q=0$ does not correspond exactly to $\beta$ equilibrium, as the latter reproduces a low $Y_Q$ that increases with $n_B$ (see values for the pure hadronic CMF model in $\beta$ equilibrium in Fig.~4 of Ref.~\cite{Yao:2023yda}, where in quark matter the $Y_Q$ is much smaller). This difference is the reason for the change in the order of the hyperons between \cref{fig:population} in this work and e.g., Fig~5 of Ref.~\cite{Roark:2018uls}, along with a small change in deconfinement chemical potential. We have addressed the effects of $\beta$ equilibrium on deconfinement and its temperature dependence in detail in Ref.~\cite{Dexheimer:2020okt}.
Furthermore, the contribution from leptons to the equation of state is actually quite small (calculated from $n_{lep}=-n_B Y_Q$) and can be ignored to first approximation as long as $Y_Q$ is small, which is the case for $\beta$ equilibrium. This can be seen for the pure hadronic CMF model in $\beta$ equilibrium in Fig.~3 of Ref.~\cite{Yao:2023yda} (in quark matter the lepton contribution is even smaller).

\begin{figure}[t!]
\centering
    \includegraphics[scale=0.35]{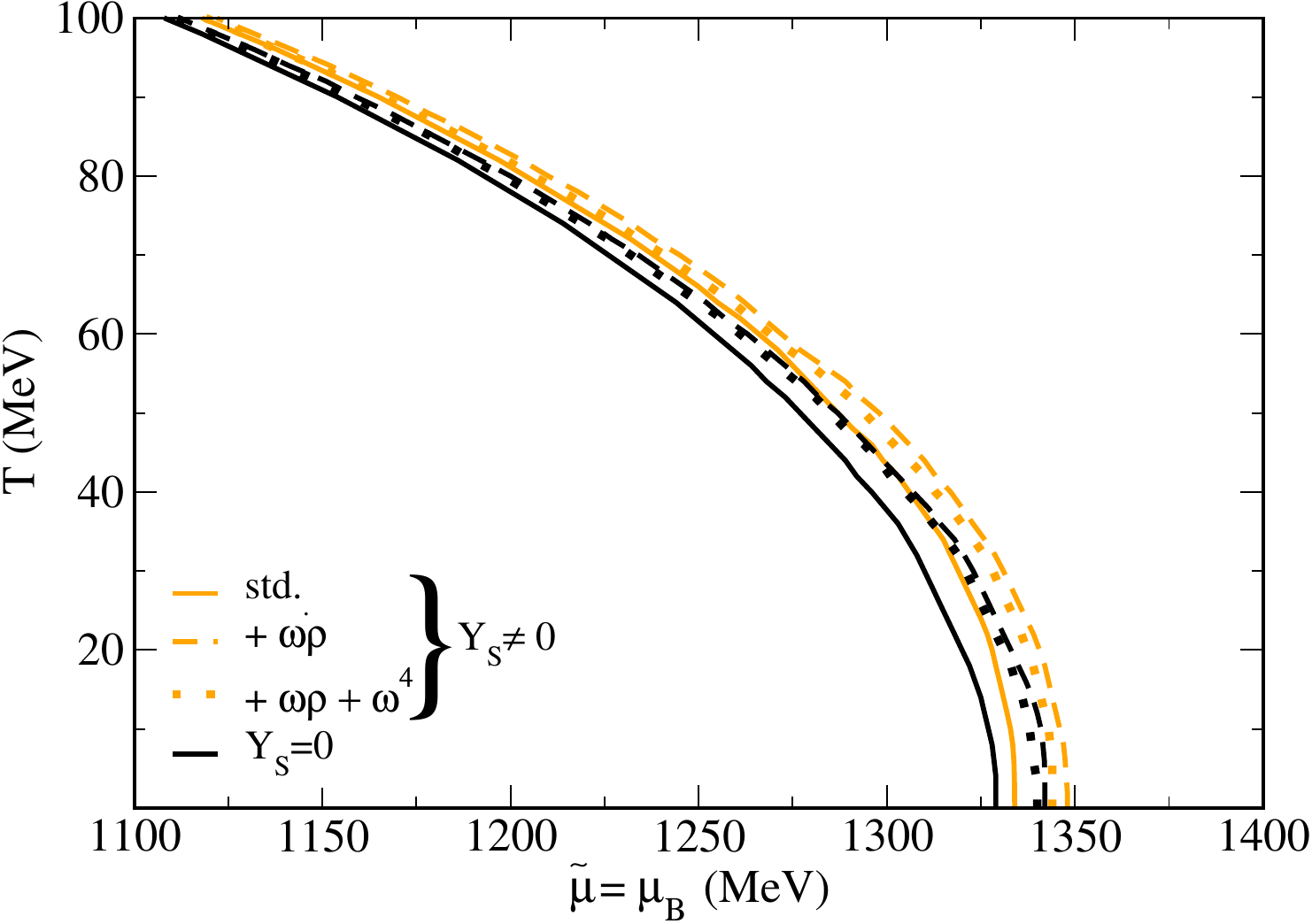}
  \caption{Temperature vs free energy (or baryon chemical potential) phase diagram for charge fraction $Y_Q=0$ matter with ExpSB~$=0.86$ shown for matter with and without zero-net strangeness and different higher-order couplings.}
  \label{fig:Tvsmub_yq0}
\end{figure}

\begin{figure}[t!]
\centering
  \includegraphics[scale=0.35]{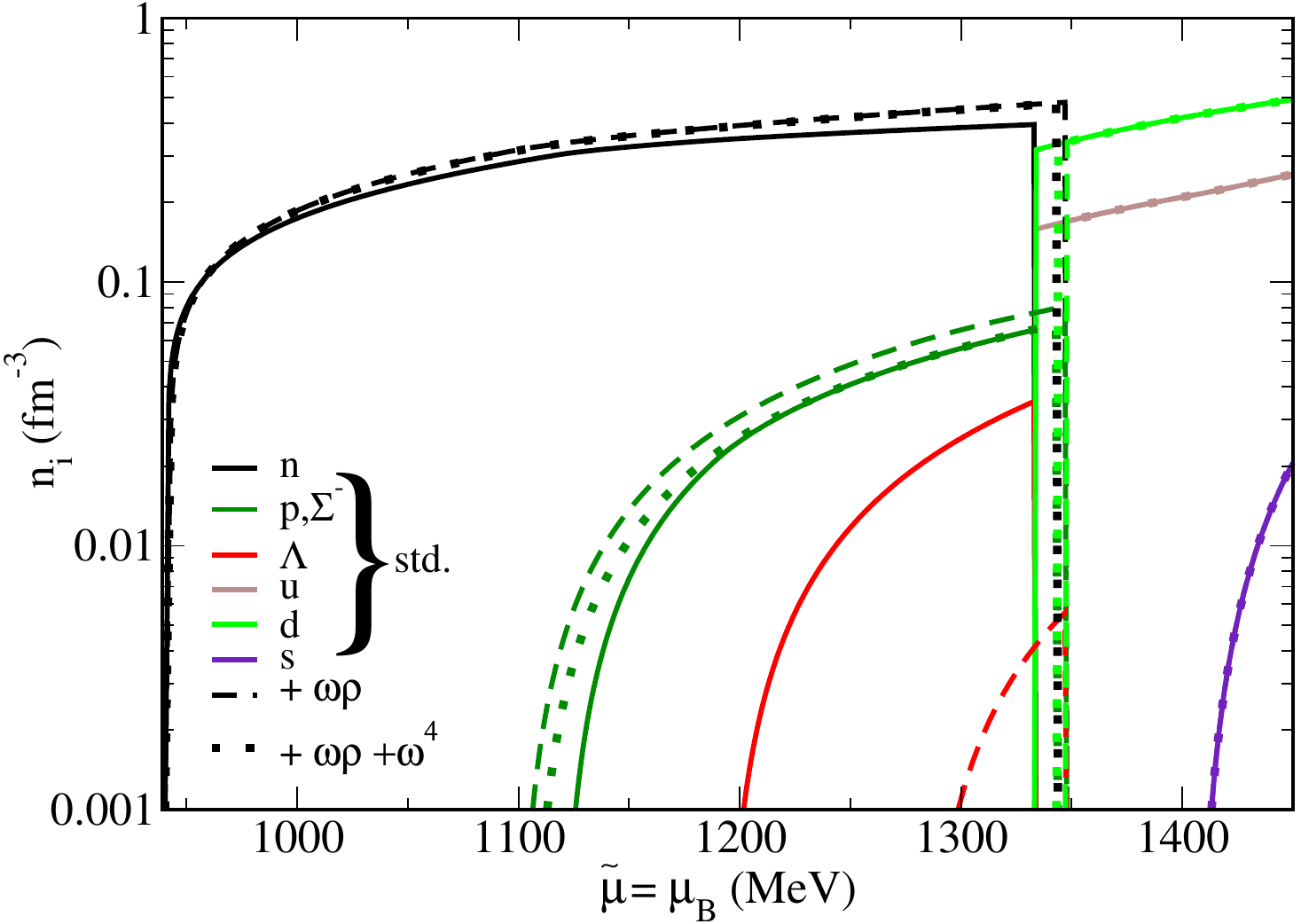}
  \caption{Particle population as a function of free energy (or baryon chemical potential) for zero-temperature, charge fraction $Y_Q=0$ matter with ExpSB~$=0.86$ and no constraints on strangeness shown for different higher-order couplings.}
  \label{fig:omrhoo4_population}
\end{figure}

As the effect of ExpSB on the phase diagram is very small (it would be even smaller for $Y_S=0$), we continue our analysis using the central value of ExpSB~$=0.86$. It reproduces $U_{\Lambda}=-28$ MeV, $U_{\Sigma}=5$ MeV, $U_{\Xi}=-18$ MeV. Now we discuss the effects of different higher-order vector couplings, $\omega\rho$ (isoscalar-isovector) and $\omega^4$ (isoscalar of maximum order that does not violate causality, see the second footnote in Ref.~\cite{Dexheimer:2020rlp}). The $\omega^4$ coupling changes the hyperon potentials for isospin-symmetric matter at saturation slightly to $U_{\Lambda}=-27$ MeV, $U_{\Sigma}=6$ MeV, $U_{\Xi}=-17$ MeV. The $\omega\rho$ coupling  does not change isospin-symmetric matter. \cref{fig:Tvsmub_yq0} shows
that the changes due to the different couplings on deconfinement are larger at lower temperatures (see different orange lines), where the coupling effects do not compete with pure thermal effects. The $\omega\rho$ interaction pushes deconfinement to a larger free energy, while $\omega^4$ has the opposite (and weaker) effect. Once more, this is due to the matter softening (stiffening) caused by the $\omega\rho$ ($\omega^4$) interactions. 

\begin{figure}[t!]
\centering
\includegraphics[scale=0.35]{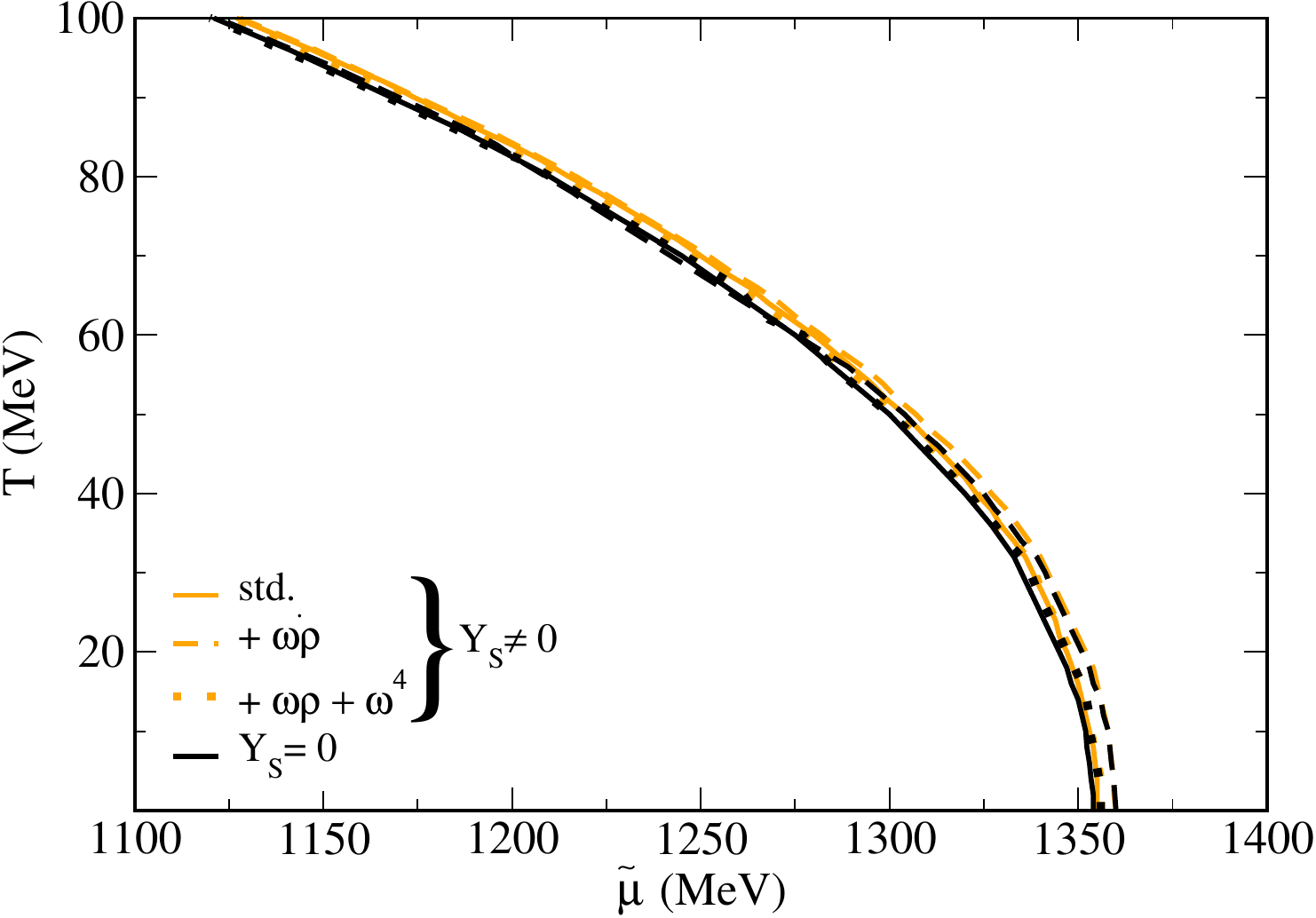}
\caption{Same as \cref{fig:Tvsmub_yq0} but for a different charge fraction, $Y_Q=0.25$.}
\label{fig:Tvsmub_025}
\end{figure}

In particular, the additional $\omega\rho$ interaction decreases the cost of producing isospin asymmetry, having the direct effect of decreasing the symmetry-energy slope $L$ from 88 to 75 MeV, thus increasing the isospin asymmetry, and consequently modifying the hyperon population at low and intermediate densities or $\mu_B$'s (see \cref{fig:omrhoo4_population}). The $\omega^4$ interaction, on the other hand, only affects the EoS of matter at large densities, but it does it directly, by contributing positively to the pressure and negatively to the energy density. Small differences at low and intermediate densities in this case are due to a slightly different $\omega$ coupling to the baryons used to fit saturation properties.

The effects of different higher-order vector couplings are also present (and slightly larger) if we constrain matter to have zero net strangeness ($Y_S=0$); see black lines in \cref{fig:Tvsmub_yq0}, in which case a non-zero strange chemical potential $\mu_S$ is introduced (see also figure with discussion in the Appendix). This zero temperature case does not allow for hyperons or strange quarks and demonstrates that the effects of different higher-order vector couplings go beyond strangeness effects. For a given coupling, allowing for net strangeness ($Y_S\neq 0$) pushes the deconfinement phase transition to large free energies and the difference is larger for larger (fixed) temperatures, when the diagram bends leftwards (see extensive discussion in Ref.~\cite{Aryal:2020ocm}).

\cref{fig:Tvsmub_025} shows that for a larger charge fraction (or, equivalently, more isospin symmetry) the pattern illustrated in \cref{fig:Tvsmub_yq0} persists, but with the differences between the couplings and different choices of strangeness being much smaller. As naturally expected, the isovector coupling is not as relevant for more isospin-symmetric matter, as well as the amount of strangeness present. Additionally, the small $\omega\rho$ difference present is now approximately balanced by the opposite change from the $\omega^4$ coupling. For even larger charge fractions (not shown here) the $\omega\rho$ interaction has no effect, only the $\omega^4$.

\begin{figure}[t!]
\centering
\includegraphics[scale=0.35]{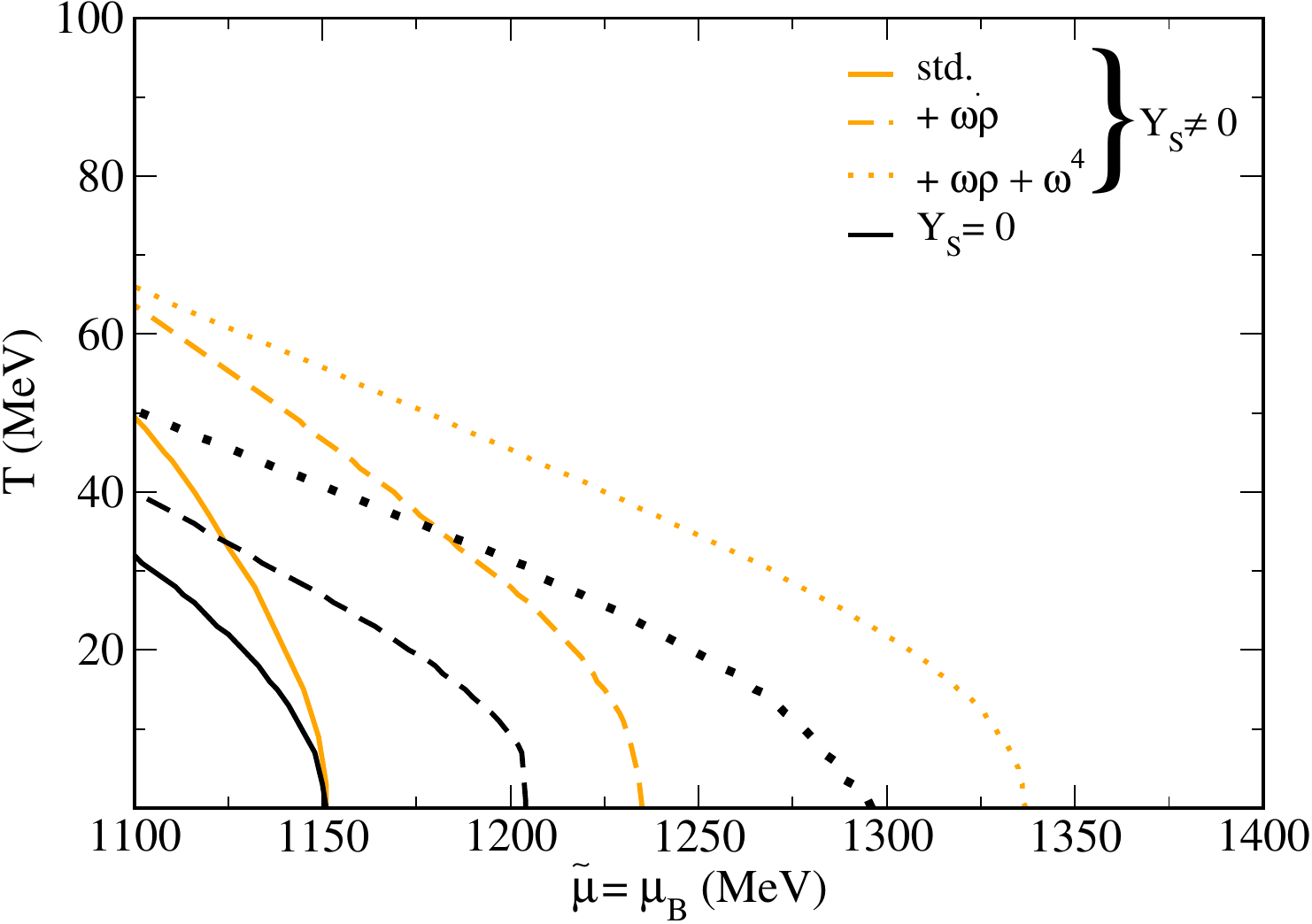}
\caption{Same as \cref{fig:Tvsmub_yq0} but for a different deconfining potential $U'_{\Phi}$.}
\label{fig:Tvsmub_newpot}
\end{figure}

We now consider the dependence of our results on the deconfining mechanism itself; we study a different deconfining potential $U'_{\Phi}$ and discuss for the first time how this affects the QCD phase diagram. We start by looking at the effects of ExpSB, which again are very small (not shown here). Then, we consider different higher-order couplings for $Y_Q=0$, for which the differences are very large. When comparing \cref{fig:Tvsmub_newpot} with \cref{fig:Tvsmub_yq0}, we immediately see that, when using $U'_{\Phi}$, different couplings or strangeness constraints generate much larger changes in the deconfining $\tilde{\mu}$. The deconfinement temperatures are also much lower for $U'_{\Phi}$. Both facts are related to $U'_{\Phi}$ producing a much weaker first-order phase transition (than using the original potential $U_{\Phi}$). The trend of having a larger deconfining $\tilde{\mu}$ for matter with net strangeness (mostly) persists. On the other hand, the order of the deconfining $\tilde{\mu}$'s for the different couplings depends on the $\Phi$ potential, with $\omega\rho$ and $\omega\rho+\omega^4$ exchanging order for most of the temperature range analyzed. 
This is due to how quark matter gets softer in the presence of the additional $\omega^4$ interaction, now pressure being lower as a function of both energy density and free energy (or baryon chemical potential). 
This correspondence is unusual but not unphysical  and it is a result of the coupling $a_1'$ being different in the case that includes $\omega^4$, which is necessary to give rise to stable hybrid stars when $\beta$ equilibrium is enforced \cite{Dexheimer:2020rlp}.

\begin{figure}[t!]
\centering
  \includegraphics[scale=0.35]{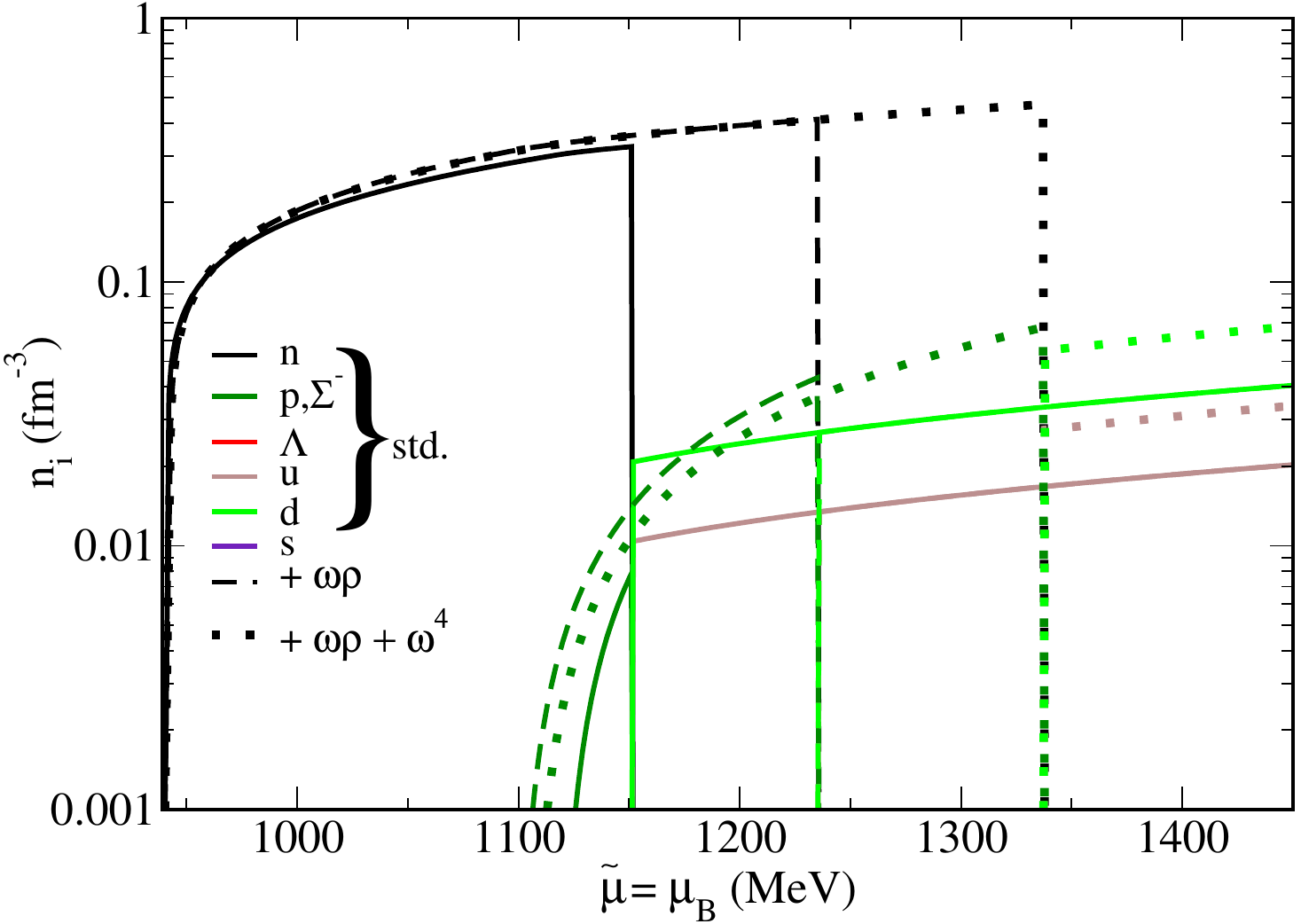}
  \caption{Same as \cref{fig:omrhoo4_population} but for a different deconfining potential $U'_{\Phi}$.}
  \label{fig:newomrhoo4_population}
\end{figure}

Looking at the particle population for zero-temperature and $Y_Q=0$, \cref{fig:newomrhoo4_population} shows that (when comparing with \cref{fig:omrhoo4_population}) the differences are not in the hadronic phase, but in the quark one. When using $U'_{\Phi}$, there are much fewer quarks and no strange quarks. The stiffening of the EoS, now presenting quark vector interactions, pushes the phase transition to lower energy densities (corresponding to lower free energies). The order of the phase transition lines also changes (as already discussed), where for $U'_{\Phi}$ the $\omega\rho+\omega^4$ the phase transition takes place at higher energy densities and free energies. 

Overall, results for the $U'_{\Phi}$ potential are much more dependent on $Y_Q$, to the point that, for large $Y_Q$'s the first-order phase transition disappears, as it becomes of higher order. To understand this, we go back to  $U_{\Phi}$, where a small change in temperature for the critical point for different $Y_Q$'s had already been pointed out \cite{Aryal:2021ojz}. In this work, we show that for $U'_{\Phi}$ this feature is exacerbated, such that the critical point goes down all the way to $T=0$. The values at which this change of phase transition order takes place are summarized in \cref{tab:YQ_summary}.

\section{Conclusion}
\label{sec:conclusion}

\begin{table}[t!]
\small
\caption{Values of charge fraction for which a first-order phase transition is found for $U'_{\Phi}$ shown for different strangeness constraints and higher-order vector couplings.}
\label{tab:YQ_summary}
\centering
\begin{tabular}{c|c|c|c}
\hline
& std. & std.+ $ \omega \rho$ & std. $ + \omega \rho + \omega^4$\\
\hline
$Y_S \neq 0$ & $Y_Q=0 - 0.14$ & $Y_Q=0 - 0.04$ & $Y_Q=0 - 0.04$ \\
\hline
$Y_S = 0$ & $Y_Q=0 - 0.15$ & $Y_Q=0 - 0.07$ & $Y_Q=0 - 0.06$ \\
\end{tabular}
\end{table}

The three pieces still missing from our understanding of dense matter are: the quantification of the softness generated by the appearance of hyperons, the quantification of the stiffness generated by isospin-asymmetric matter, and how/where deconfinement to quark matter takes place. In this work, we considered all these pieces, using one consistent description to study how couplings that directly affect hyperon potentials (and their appearance in dense matter) and how vector interactions that directly affect isospin can change the free energy/temperature and strength of deconfinement to quark matter. We do not assume weak ($\beta$) equilibrium but vary the charge fraction and also study different possibilities for strangeness (either constrained to be zero or not).

We find that hyperon potentials $U_H$ varied within or near the experimental range (of $\lesssim 10$ MeV each) do not significantly affect deconfinement  at any temperature, but higher-order interactions that modify the isospin sensitivity of the density, $L$, by $7$ MeV do. The changes are more significant for low temperature, non-constrained strangeness, and lower charge fraction $Y_Q$ (or very isospin-asymmetric matter), pushing the deconfinement coexistence free energy by up to $15$ MeV  (corresponding to a shift in density for $Y_Q=0$ at $T=0$ of $\sim0.5$ $n_{\rm{sat}}$). These conditions are  similar to the ones reproduced by neutron stars.
To verify the consistency of our results, we also investigate a different deconfining potential $U'(\Phi)$, which has a weaker dependence on baryon chemical potential and reproduces, as a consequence, a weaker first-order phase transition. As a result,  there is still no significant effect of the parameter (ExpSB) that controls the hyperon potentials on deconfinement but, using $U'(\Phi)$, higher-order vector interactions (that modify $L$) can increase the deconfinement coexistence free energy much more  than $U(\Phi)$, up to $185$ MeV (corresponding to a shift in density for $Y_Q=0$ at $T=0$ of $\sim 2$ $n_{\rm{sat}}$).

Together, these results point to the importance of understanding recent controversies surrounding the interpretation of PREXII results, especially for astrophysics. While we were finalizing this work, Ref.~\cite{Lopes:2023dnx} appeared online. There, the authors study the effect of $L$ on $\beta$-equilibrated matter at $T=0$ using a modified Walecka model combined with the vector-bag model. Their results point to lower $L$ values being associated with a higher critical baryon chemical potential for the phase transition (in agreement with our results), thereby decreasing the quark content and increasing the maximum mass of neutron stars.  To the best of our knowledge, this is the only other work in the literature addressing this relationship.

We acknowledges support from the National Science Foundation
under grants PHY1748621, MUSES OAC-2103680, and
NP3M PHY-2116686.

\section*{Appendix}

As discussed in \cref{sec:RnD}, constraining matter (in our case to zero-net strangeness, $Y_S$=0) gives rise to a non-zero $\mu_S$. In \cref{cps} we show $\mu_S$ as a function of free energy for two chosen temperatures for the cases corresponding to the black lines in
\cref{fig:Tvsmub_yq0,fig:Tvsmub_025,fig:Tvsmub_newpot}.
In every scenario, the absolute value of $\mu_S$ increases with increasing $\mu_B$, and falls at (or around for large temperatures close to the critical point) the phase transition. The cost to control strangeness is highest for $U_\Phi$ with $Y_Q=0$, decreasing for $U_\Phi$ with $Y_Q=0.25$, and is lowest for $U'_\Phi$ with $Y_Q=0$. The difference in the hadronic phase behavior between $U_\Phi$ and $U'_\Phi$ is due to quarks appearing in the hadronic phase at large $T$ \cite{Dexheimer:2009hi}.

\renewcommand\thefigure{A\arabic{figure}} 
\setcounter{figure}{0}  
\begin{figure}[t!]
\centering
  \includegraphics[scale=0.35]{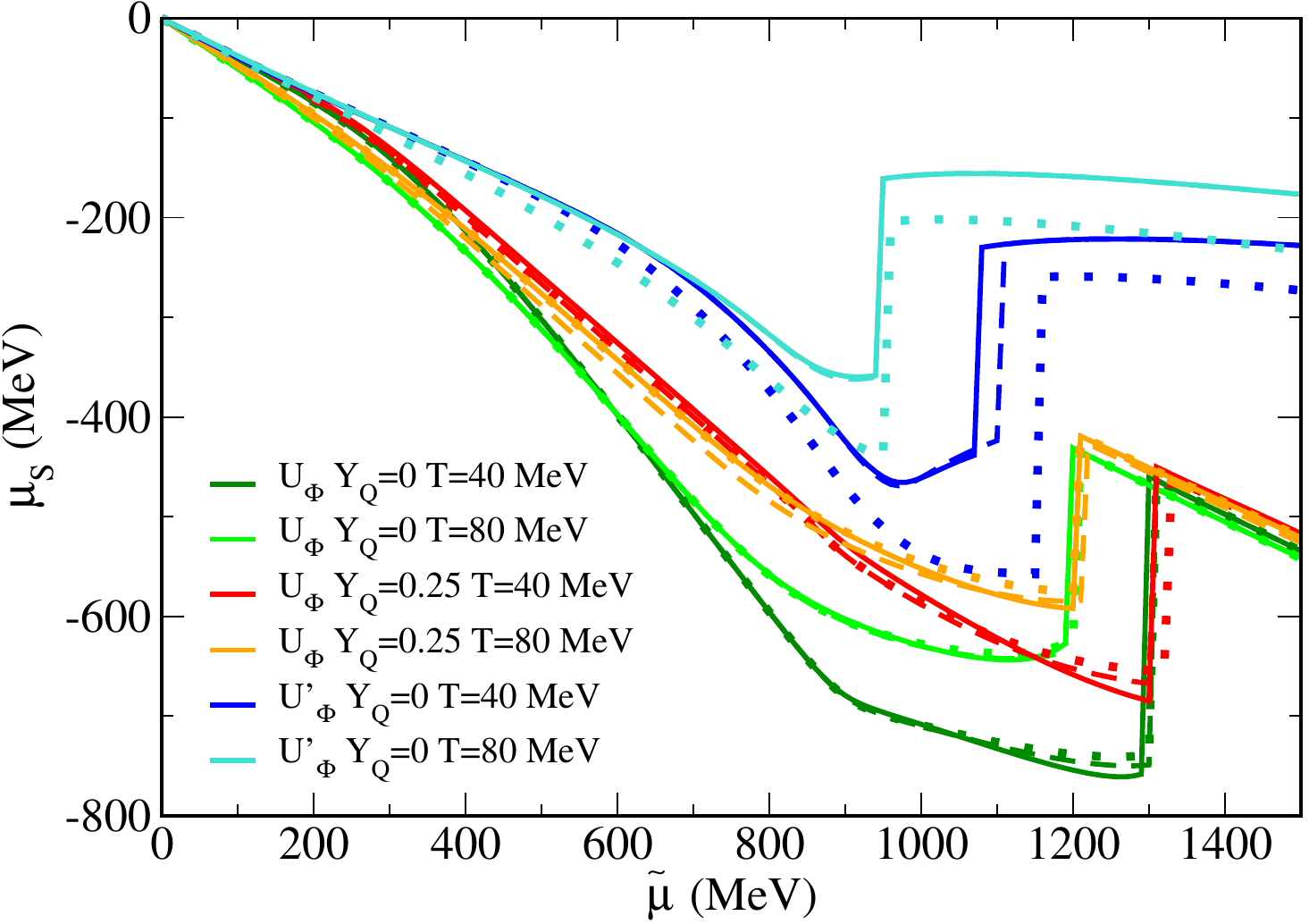}
  \caption{Strange chemical potential as a function of free energy for all cases discussed. Full lines show standard higher-order interactions, while dashed also include $\omega\rho$ and dotted lines $\omega\rho$ and $\omega^4$ terms.}
  \label{cps}
\end{figure}

\bibliographystyle{elsarticle-num}

\bibliography{inspire,Not_inspire}

\begin{thebibliography}{10}
\expandafter\ifx\csname url\endcsname\relax
  \def\url#1{\texttt{#1}}\fi
\expandafter\ifx\csname urlprefix\endcsname\relax\def\urlprefix{URL }\fi
\expandafter\ifx\csname href\endcsname\relax
  \def\href#1#2{#2} \def\path#1{#1}\fi

\bibitem{PREX:2021umo}
D.~Adhikari, et~al., {Accurate Determination of the Neutron Skin Thickness of $^{208}$Pb through Parity-Violation in Electron Scattering}, Phys. Rev. Lett. 126~(17) (2021) 172502.
\newblock \href {http://arxiv.org/abs/2102.10767} {\path{arXiv:2102.10767}}, \href {https://doi.org/10.1103/PhysRevLett.126.172502} {\path{doi:10.1103/PhysRevLett.126.172502}}.

\bibitem{Furnstahl:2001un}
R.~J. Furnstahl, {Neutron radii in mean field models}, Nucl. Phys. A 706 (2002) 85--110.
\newblock \href {http://arxiv.org/abs/nucl-th/0112085} {\path{arXiv:nucl-th/0112085}}, \href {https://doi.org/10.1016/S0375-9474(02)00867-9} {\path{doi:10.1016/S0375-9474(02)00867-9}}.

\bibitem{Chen:2010qx}
L.-W. Chen, C.~M. Ko, B.-A. Li, J.~Xu, {Density slope of the nuclear symmetry energy from the neutron skin thickness of heavy nuclei}, Phys. Rev. C 82 (2010) 024321.
\newblock \href {http://arxiv.org/abs/1004.4672} {\path{arXiv:1004.4672}}, \href {https://doi.org/10.1103/PhysRevC.82.024321} {\path{doi:10.1103/PhysRevC.82.024321}}.

\bibitem{Reed:2021nqk}
B.~T. Reed, F.~J. Fattoyev, C.~J. Horowitz, J.~Piekarewicz, {Implications of PREX-2 on the Equation of State of Neutron-Rich Matter}, Phys. Rev. Lett. 126~(17) (2021) 172503.
\newblock \href {http://arxiv.org/abs/2101.03193} {\path{arXiv:2101.03193}}, \href {https://doi.org/10.1103/PhysRevLett.126.172503} {\path{doi:10.1103/PhysRevLett.126.172503}}.

\bibitem{Reinhard:2021utv}
P.-G. Reinhard, X.~Roca-Maza, W.~Nazarewicz, {Information Content of the Parity-Violating Asymmetry in Pb208}, Phys. Rev. Lett. 127~(23) (2021) 232501.
\newblock \href {http://arxiv.org/abs/2105.15050} {\path{arXiv:2105.15050}}, \href {https://doi.org/10.1103/PhysRevLett.127.232501} {\path{doi:10.1103/PhysRevLett.127.232501}}.

\bibitem{Zenihiro:2010zz}
J.~Zenihiro, et~al., {Neutron density distributions of Pb-204, Pb-206, Pb-208 deduced via proton elastic scattering at Ep=295 MeV}, Phys. Rev. C 82 (2010) 044611.
\newblock \href {https://doi.org/10.1103/PhysRevC.82.044611} {\path{doi:10.1103/PhysRevC.82.044611}}.

\bibitem{Giacalone:2023cet}
G.~Giacalone, G.~Nijs, W.~van~der Schee, {Determination of the Neutron Skin of Pb208 from Ultrarelativistic Nuclear Collisions}, Phys. Rev. Lett. 131~(20) (2023) 202302.
\newblock \href {http://arxiv.org/abs/2305.00015} {\path{arXiv:2305.00015}}, \href {https://doi.org/10.1103/PhysRevLett.131.202302} {\path{doi:10.1103/PhysRevLett.131.202302}}.

\bibitem{Li:2019xxz}
B.-A. Li, P.~G. Krastev, D.-H. Wen, N.-B. Zhang, {Towards Understanding Astrophysical Effects of Nuclear Symmetry Energy}, Eur. Phys. J. A 55~(7) (2019) 117.
\newblock \href {http://arxiv.org/abs/1905.13175} {\path{arXiv:1905.13175}}, \href {https://doi.org/10.1140/epja/i2019-12780-8} {\path{doi:10.1140/epja/i2019-12780-8}}.

\bibitem{LIGOScientific:2018hze}
B.~P. Abbott, et~al., {Properties of the binary neutron star merger GW170817}, Phys. Rev. X 9~(1) (2019) 011001.
\newblock \href {http://arxiv.org/abs/1805.11579} {\path{arXiv:1805.11579}}, \href {https://doi.org/10.1103/PhysRevX.9.011001} {\path{doi:10.1103/PhysRevX.9.011001}}.

\bibitem{Millener:1988hp}
D.~J. Millener, C.~B. Dover, A.~Gal, {Lambda Nucleus Single Particle Potentials}, Phys. Rev. C 38 (1988) 2700--2708.
\newblock \href {https://doi.org/10.1103/PhysRevC.38.2700} {\path{doi:10.1103/PhysRevC.38.2700}}.

\bibitem{Fortin:2017dsj}
M.~Fortin, M.~Oertel, C.~Provid\^encia, {Hyperons in hot dense matter: what do the constraints tell us for equation of state?}, Publ. Astron. Soc. Austral. 35 (2018) 44.
\newblock \href {http://arxiv.org/abs/1711.09427} {\path{arXiv:1711.09427}}, \href {https://doi.org/10.1017/pasa.2018.32} {\path{doi:10.1017/pasa.2018.32}}.

\bibitem{Gal:2016boi}
A.~Gal, E.~V. Hungerford, D.~J. Millener, {Strangeness in nuclear physics}, Rev. Mod. Phys. 88~(3) (2016) 035004.
\newblock \href {http://arxiv.org/abs/1605.00557} {\path{arXiv:1605.00557}}, \href {https://doi.org/10.1103/RevModPhys.88.035004} {\path{doi:10.1103/RevModPhys.88.035004}}.

\bibitem{AGSE885:1999erv}
P.~Khaustov, et~al., {Evidence of Xi hypernuclear production in the C-12(K-,K+)(Xi)Be-12 reaction}, Phys. Rev. C 61 (2000) 054603.
\newblock \href {http://arxiv.org/abs/nucl-ex/9912007} {\path{arXiv:nucl-ex/9912007}}, \href {https://doi.org/10.1103/PhysRevC.61.054603} {\path{doi:10.1103/PhysRevC.61.054603}}.

\bibitem{Fabbietti:2020bfg}
L.~Fabbietti, V.~Mantovani~Sarti, O.~Vazquez~Doce, {Study of the Strong Interaction Among Hadrons with Correlations at the LHC}, Ann. Rev. Nucl. Part. Sci. 71 (2021) 377--402.
\newblock \href {http://arxiv.org/abs/2012.09806} {\path{arXiv:2012.09806}}, \href {https://doi.org/10.1146/annurev-nucl-102419-034438} {\path{doi:10.1146/annurev-nucl-102419-034438}}.

\bibitem{ALICE:2020mfd}
A.~Collaboration, et~al., {Unveiling the strong interaction among hadrons at the LHC}, Nature 588 (2020) 232--238, [Erratum: Nature 590, E13 (2021)].
\newblock \href {http://arxiv.org/abs/2005.11495} {\path{arXiv:2005.11495}}, \href {https://doi.org/10.1038/s41586-020-3001-6} {\path{doi:10.1038/s41586-020-3001-6}}.

\bibitem{Inoue:2019jme}
T.~Inoue, {Hyperon Forces from QCD and Their Applications}, JPS Conf. Proc. 26 (2019) 023018.
\newblock \href {https://doi.org/10.7566/JPSCP.26.023018} {\path{doi:10.7566/JPSCP.26.023018}}.

\bibitem{Iida:1998pi}
K.~Iida, K.~Sato, {Effects of hyperons on the dynamical deconfinement transition in cold neutron star matter}, Phys. Rev. C 58 (1998) 2538--2559.
\newblock \href {http://arxiv.org/abs/nucl-th/9808056} {\path{arXiv:nucl-th/9808056}}, \href {https://doi.org/10.1103/PhysRevC.58.2538} {\path{doi:10.1103/PhysRevC.58.2538}}.

\bibitem{Carroll:2008sv}
J.~D. Carroll, D.~B. Leinweber, A.~G. Williams, A.~W. Thomas, {Phase Transition from QMC Hyperonic Matter to Deconfined Quark Matter}, Phys. Rev. C 79 (2009) 045810.
\newblock \href {http://arxiv.org/abs/0809.0168} {\path{arXiv:0809.0168}}, \href {https://doi.org/10.1103/PhysRevC.79.045810} {\path{doi:10.1103/PhysRevC.79.045810}}.

\bibitem{Zschiesche:1999gf}
D.~Zschiesche, P.~Papazoglou, C.~W. Beckmann, S.~Schramm, J.~Schaffner-Bielich, H.~Stoecker, W.~Greiner, {Chiral model for dense, hot and strange hadronic matter}, Nucl. Phys. A 663 (2000) 737--740.
\newblock \href {http://arxiv.org/abs/nucl-th/9908072} {\path{arXiv:nucl-th/9908072}}, \href {https://doi.org/10.1016/S0375-9474(99)00707-1} {\path{doi:10.1016/S0375-9474(99)00707-1}}.

\bibitem{Papazoglou:1997uw}
P.~Papazoglou, S.~Schramm, J.~Schaffner-Bielich, H.~Stoecker, W.~Greiner, {Chiral Lagrangian for strange hadronic matter}, Phys. Rev. C 57 (1998) 2576--2588.
\newblock \href {http://arxiv.org/abs/nucl-th/9706024} {\path{arXiv:nucl-th/9706024}}, \href {https://doi.org/10.1103/PhysRevC.57.2576} {\path{doi:10.1103/PhysRevC.57.2576}}.

\bibitem{Wang:2001hw}
P.~Wang, Z.~Y. Zhang, Y.~W. Yu, R.~K. Su, Q.~Song, {Strange hadronic matter in a chiral SU(3) quark mean field model}, Nucl. Phys. A 688 (2001) 791--807.
\newblock \href {https://doi.org/10.1016/S0375-9474(00)00580-7} {\path{doi:10.1016/S0375-9474(00)00580-7}}.

\bibitem{Weissenborn:2011kb}
S.~Weissenborn, D.~Chatterjee, J.~Schaffner-Bielich, {Hyperons and massive neutron stars: the role of hyperon potentials}, Nucl. Phys. A 881 (2012) 62--77.
\newblock \href {http://arxiv.org/abs/1111.6049} {\path{arXiv:1111.6049}}, \href {https://doi.org/10.1016/j.nuclphysa.2012.02.012} {\path{doi:10.1016/j.nuclphysa.2012.02.012}}.

\bibitem{Petschauer:2015nea}
S.~Petschauer, J.~Haidenbauer, N.~Kaiser, U.-G. Mei\ss{}ner, W.~Weise, {Hyperons in nuclear matter from SU(3) chiral effective field theory}, Eur. Phys. J. A 52~(1) (2016) 15.
\newblock \href {http://arxiv.org/abs/1507.08808} {\path{arXiv:1507.08808}}, \href {https://doi.org/10.1140/epja/i2016-16015-4} {\path{doi:10.1140/epja/i2016-16015-4}}.

\bibitem{Wang:2002pza}
P.~Wang, V.~E. Lyubovitskij, T.~Gutsche, A.~Faessler, {Strange quark matter in a chiral SU(3) quark mean field model}, Phys. Rev. C 67 (2003) 015210.
\newblock \href {http://arxiv.org/abs/hep-ph/0205251} {\path{arXiv:hep-ph/0205251}}, \href {https://doi.org/10.1103/PhysRevC.67.015210} {\path{doi:10.1103/PhysRevC.67.015210}}.

\bibitem{Kumari:2020mci}
M.~Kumari, A.~Kumar, {Quark Matter within Polyakov Chiral SU(3) Quark Mean Field Model at Finite Temperature}, Eur. Phys. J. Plus 136~(1) (2021) 19.
\newblock \href {http://arxiv.org/abs/2003.12780} {\path{arXiv:2003.12780}}, \href {https://doi.org/10.1140/epjp/s13360-020-00999-0} {\path{doi:10.1140/epjp/s13360-020-00999-0}}.

\bibitem{Fortin:2020qin}
M.~Fortin, A.~R. Raduta, S.~Avancini, C.~Provid\^encia, {Relativistic hypernuclear compact stars with calibrated equations of state}, Phys. Rev. D 101~(3) (2020) 034017.
\newblock \href {http://arxiv.org/abs/2001.08036} {\path{arXiv:2001.08036}}, \href {https://doi.org/10.1103/PhysRevD.101.034017} {\path{doi:10.1103/PhysRevD.101.034017}}.

\bibitem{Schaffner-Bielich:2000igu}
J.~Schaffner-Bielich, A.~Gal, {Properties of strange hadronic matter in bulk and in finite systems}, Phys. Rev. C 62 (2000) 034311.
\newblock \href {http://arxiv.org/abs/nucl-th/0005060} {\path{arXiv:nucl-th/0005060}}, \href {https://doi.org/10.1103/PhysRevC.62.034311} {\path{doi:10.1103/PhysRevC.62.034311}}.

\bibitem{Hu:2021ket}
J.~Hu, Y.~Zhang, H.~Shen, {The $\Xi$-nuclear potential constrained by recent $\Xi^-$ hypernuclei experiments}, J. Phys. G 49~(2) (2022) 025104.
\newblock \href {http://arxiv.org/abs/2104.13567} {\path{arXiv:2104.13567}}, \href {https://doi.org/10.1088/1361-6471/ac4469} {\path{doi:10.1088/1361-6471/ac4469}}.

\bibitem{Fu:2022eeb}
H.~R. Fu, J.~J. Li, A.~Sedrakian, F.~Weber, {Massive relativistic compact stars from SU(3) symmetric quark models}, Phys. Lett. B 834 (2022) 137470.
\newblock \href {http://arxiv.org/abs/2209.05699} {\path{arXiv:2209.05699}}, \href {https://doi.org/10.1016/j.physletb.2022.137470} {\path{doi:10.1016/j.physletb.2022.137470}}.

\bibitem{Kochankovski:2023trc}
H.~Kochankovski, A.~Ramos, L.~Tolos, {Hyperonic Uncertainties in Neutron Stars, Mergers and Supernovae} (9 2023).
\newblock \href {http://arxiv.org/abs/2309.14879} {\path{arXiv:2309.14879}}.

\bibitem{Gusakov:2014ota}
M.~E. Gusakov, P.~Haensel, E.~M. Kantor, {Physics input for modelling superfluid neutron stars with hyperon cores}, Mon. Not. Roy. Astron. Soc. 439~(1) (2014) 318--333.
\newblock \href {http://arxiv.org/abs/1401.2827} {\path{arXiv:1401.2827}}, \href {https://doi.org/10.1093/mnras/stt2438} {\path{doi:10.1093/mnras/stt2438}}.

\bibitem{Chatterjee:2015pua}
D.~Chatterjee, I.~Vida\~na, {Do hyperons exist in the interior of neutron stars?}, Eur. Phys. J. A 52~(2) (2016) 29.
\newblock \href {http://arxiv.org/abs/1510.06306} {\path{arXiv:1510.06306}}, \href {https://doi.org/10.1140/epja/i2016-16029-x} {\path{doi:10.1140/epja/i2016-16029-x}}.

\bibitem{Isaka:2017nuc}
M.~Isaka, Y.~Yamamoto, T.~A. Rijken, {Effects of hyperonic many-body force on $B_\Lambda$ values of hypernuclei}, Phys. Rev. C 95~(4) (2017) 044308.
\newblock \href {http://arxiv.org/abs/1703.03117} {\path{arXiv:1703.03117}}, \href {https://doi.org/10.1103/PhysRevC.95.044308} {\path{doi:10.1103/PhysRevC.95.044308}}.

\bibitem{Zhao:2017nlw}
X.-F. Zhao, {The hyperons in the massive neutron star PSR J0348+0432} (12 2017).
\newblock \href {http://arxiv.org/abs/1712.08854} {\path{arXiv:1712.08854}}, \href {https://doi.org/10.6122/CJP.20150601D} {\path{doi:10.6122/CJP.20150601D}}.

\bibitem{Djapo:2008au}
H.~Djapo, B.-J. Schaefer, J.~Wambach, {On the appearance of hyperons in neutron stars}, Phys. Rev. C 81 (2010) 035803.
\newblock \href {http://arxiv.org/abs/0811.2939} {\path{arXiv:0811.2939}}, \href {https://doi.org/10.1103/PhysRevC.81.035803} {\path{doi:10.1103/PhysRevC.81.035803}}.

\bibitem{Miyatsu:2022wuy}
T.~Miyatsu, M.-K. Cheoun, K.~Saito, {Asymmetric Nuclear Matter in Relativistic Mean-field Models with Isoscalar- and Isovector-meson Mixing}, Astrophys. J. 929~(1) (2022) 82.
\newblock \href {http://arxiv.org/abs/2202.06468} {\path{arXiv:2202.06468}}, \href {https://doi.org/10.3847/1538-4357/ac5f40} {\path{doi:10.3847/1538-4357/ac5f40}}.

\bibitem{Lopes:2023dnx}
L.~L. Lopes, D.~P. Menezes, M.~R. Pelicer, {Correlation between the symmetry energy slope and the deconfinement phase transition} (10 2023).
\newblock \href {http://arxiv.org/abs/2311.00125} {\path{arXiv:2311.00125}}.

\bibitem{Drischler:2013iza}
C.~Drischler, V.~Soma, A.~Schwenk, {Microscopic calculations and energy expansions for neutron-rich matter}, Phys. Rev. C 89~(2) (2014) 025806.
\newblock \href {http://arxiv.org/abs/1310.5627} {\path{arXiv:1310.5627}}, \href {https://doi.org/10.1103/PhysRevC.89.025806} {\path{doi:10.1103/PhysRevC.89.025806}}.

\bibitem{Horowitz:2014bja}
C.~J. Horowitz, E.~F. Brown, Y.~Kim, W.~G. Lynch, R.~Michaels, A.~Ono, J.~Piekarewicz, M.~B. Tsang, H.~H. Wolter, {A way forward in the study of the symmetry energy: experiment, theory, and observation}, J. Phys. G 41 (2014) 093001.
\newblock \href {http://arxiv.org/abs/1401.5839} {\path{arXiv:1401.5839}}, \href {https://doi.org/10.1088/0954-3899/41/9/093001} {\path{doi:10.1088/0954-3899/41/9/093001}}.

\bibitem{Trautmann:2017xlx}
W.~Trautmann, H.~H. Wolter, {Elliptic Flow and the Nuclear Equation of State} (2018).
\newblock \href {http://arxiv.org/abs/1712.03093} {\path{arXiv:1712.03093}}, \href {https://doi.org/10.1142/9789813234284_0023} {\path{doi:10.1142/9789813234284_0023}}.

\bibitem{Sotani:2015lya}
H.~Sotani, K.~Iida, K.~Oyamatsu, {Constraining the density dependence of the nuclear symmetry energy from an X-ray bursting neutron star}, Phys. Rev. C 91~(1) (2015) 015805.
\newblock \href {http://arxiv.org/abs/1501.01698} {\path{arXiv:1501.01698}}, \href {https://doi.org/10.1103/PhysRevC.91.015805} {\path{doi:10.1103/PhysRevC.91.015805}}.

\bibitem{Providencia:2013dsa}
C.~Provid\^encia, S.~S. Avancini, R.~Cavagnoli, S.~Chiacchiera, C.~Ducoin, F.~Grill, J.~Margueron, D.~P. Menezes, A.~Rabhi, I.~Vida\~na, {Imprint of the symmetry energy on the inner crust and strangeness content of neutron stars}, Eur. Phys. J. A 50 (2014) 44.
\newblock \href {http://arxiv.org/abs/1307.1436} {\path{arXiv:1307.1436}}, \href {https://doi.org/10.1140/epja/i2014-14044-7} {\path{doi:10.1140/epja/i2014-14044-7}}.

\bibitem{Jiang:2012zs}
W.-Z. Jiang, R.-Y. Yang, D.-R. Zhang, {Symmetry energy softening in nuclear matter with non-nucleonic constituents}, Phys. Rev. C 87~(6) (2013) 064314.
\newblock \href {http://arxiv.org/abs/1212.3686} {\path{arXiv:1212.3686}}, \href {https://doi.org/10.1103/PhysRevC.87.064314} {\path{doi:10.1103/PhysRevC.87.064314}}.

\bibitem{Steiner:2004fi}
A.~W. Steiner, M.~Prakash, J.~M. Lattimer, P.~J. Ellis, {Isospin asymmetry in nuclei and neutron stars}, Phys. Rept. 411 (2005) 325--375.
\newblock \href {http://arxiv.org/abs/nucl-th/0410066} {\path{arXiv:nucl-th/0410066}}, \href {https://doi.org/10.1016/j.physrep.2005.02.004} {\path{doi:10.1016/j.physrep.2005.02.004}}.

\bibitem{Millerson:2019jkg}
R.~Millerson, F.~Sammarruca, {Properties of isospin asymmetric matter derived from chiral effective field theory} (6 2019).
\newblock \href {http://arxiv.org/abs/1906.02905} {\path{arXiv:1906.02905}}.

\bibitem{Negreiros:2018cho}
R.~Negreiros, L.~Tolos, M.~Centelles, A.~Ramos, V.~Dexheimer, {Cooling of Small and Massive Hyperonic Stars}, Astrophys. J. 863~(1) (2018) 104.
\newblock \href {http://arxiv.org/abs/1804.00334} {\path{arXiv:1804.00334}}, \href {https://doi.org/10.3847/1538-4357/aad049} {\path{doi:10.3847/1538-4357/aad049}}.

\bibitem{Li:2020ass}
B.-A. Li, M.~Magno, {Curvature-slope correlation of nuclear symmetry energy and its imprints on the crust-core transition, radius and tidal deformability of canonical neutron stars}, Phys. Rev. C 102~(4) (2020) 045807.
\newblock \href {http://arxiv.org/abs/2008.11338} {\path{arXiv:2008.11338}}, \href {https://doi.org/10.1103/PhysRevC.102.045807} {\path{doi:10.1103/PhysRevC.102.045807}}.

\bibitem{Most:2021ktk}
E.~R. Most, C.~A. Raithel, {Impact of the nuclear symmetry energy on the post-merger phase of a binary neutron star coalescence}, Phys. Rev. D 104~(12) (2021) 124012.
\newblock \href {http://arxiv.org/abs/2107.06804} {\path{arXiv:2107.06804}}, \href {https://doi.org/10.1103/PhysRevD.104.124012} {\path{doi:10.1103/PhysRevD.104.124012}}.

\bibitem{Yang:2023ogo}
Y.~Yang, M.~Hippert, E.~Speranza, J.~Noronha, {Far-from-equilibrium bulk-viscous transport coefficients in neutron star mergers} (9 2023).
\newblock \href {http://arxiv.org/abs/2309.01864} {\path{arXiv:2309.01864}}.

\bibitem{Hutauruk:2023mjj}
P.~T.~P. Hutauruk, H.~Gil, S.-i. Nam, C.~H. Hyun, {Effects of Symmetry Energy on the Equation of State for Hybrid Neutron Stars} (7 2023).
\newblock \href {http://arxiv.org/abs/2307.09038} {\path{arXiv:2307.09038}}.

\bibitem{Krastev:2021reh}
P.~G. Krastev, {Translating Neutron Star Observations to Nuclear Symmetry Energy via Deep Neural Networks}, Galaxies 10~(1) (2022) 16.
\newblock \href {http://arxiv.org/abs/2112.04089} {\path{arXiv:2112.04089}}, \href {https://doi.org/10.3390/galaxies10010016} {\path{doi:10.3390/galaxies10010016}}.

\bibitem{Thapa:2021syu}
V.~B. Thapa, M.~Sinha, {Influence of the nuclear symmetry energy slope on observables of compact stars with $\Delta$-admixed hypernuclear matter}, Phys. Rev. C 105~(1) (2022) 015802.
\newblock \href {http://arxiv.org/abs/2112.12629} {\path{arXiv:2112.12629}}, \href {https://doi.org/10.1103/PhysRevC.105.015802} {\path{doi:10.1103/PhysRevC.105.015802}}.

\bibitem{Ghosh:2022lam}
S.~Ghosh, B.~K. Pradhan, D.~Chatterjee, J.~Schaffner-Bielich, {Multi-Physics Constraints at Different Densities to Probe Nuclear Symmetry Energy in Hyperonic Neutron Stars}, Front. Astron. Space Sci. 9 (2022) 864294.
\newblock \href {http://arxiv.org/abs/2203.03156} {\path{arXiv:2203.03156}}, \href {https://doi.org/10.3389/fspas.2022.864294} {\path{doi:10.3389/fspas.2022.864294}}.

\bibitem{Suzuki:2022mow}
T.~Suzuki, {The relationship of the neutron skin thickness to the symmetry energy and its slope}, PTEP 2022~(6) (2022) 063D01.
\newblock \href {http://arxiv.org/abs/2204.11013} {\path{arXiv:2204.11013}}, \href {https://doi.org/10.1093/ptep/ptac083} {\path{doi:10.1093/ptep/ptac083}}.

\bibitem{Char:2023fue}
P.~Char, C.~Mondal, F.~Gulminelli, M.~Oertel, {Generalised description of Neutron Star matter with nucleonic Relativistic Density Functional} (7 2023).
\newblock \href {http://arxiv.org/abs/2307.12364} {\path{arXiv:2307.12364}}.

\bibitem{Tsang:2023vhh}
C.~Y. Tsang, M.~B. Tsang, W.~G. Lynch, R.~Kumar, C.~J. Horowitz, {Determination of the Equation of State from Nuclear Experiments and Neutron Star Observations} (10 2023).
\newblock \href {http://arxiv.org/abs/2310.11588} {\path{arXiv:2310.11588}}.

\bibitem{Lim:2019som}
Y.~Lim, J.~W. Holt, {Bayesian modeling of the nuclear equation of state for neutron star tidal deformabilities and GW170817}, Eur. Phys. J. A 55~(11) (2019) 209.
\newblock \href {http://arxiv.org/abs/1902.05502} {\path{arXiv:1902.05502}}, \href {https://doi.org/10.1140/epja/i2019-12917-9} {\path{doi:10.1140/epja/i2019-12917-9}}.

\bibitem{Chen:2015gba}
L.-W. Chen, {Symmetry energy systematics and its high density behavior}, EPJ Web Conf. 88 (2015) 00017.
\newblock \href {http://arxiv.org/abs/1506.09057} {\path{arXiv:1506.09057}}, \href {https://doi.org/10.1051/epjconf/20158800017} {\path{doi:10.1051/epjconf/20158800017}}.

\bibitem{Dexheimer:2009hi}
V.~A. Dexheimer, S.~Schramm, {A Novel Approach to Model Hybrid Stars}, Phys. Rev. C 81 (2010) 045201.
\newblock \href {http://arxiv.org/abs/0901.1748} {\path{arXiv:0901.1748}}, \href {https://doi.org/10.1103/PhysRevC.81.045201} {\path{doi:10.1103/PhysRevC.81.045201}}.

\bibitem{Horowitz:2002mb}
C.~J. Horowitz, J.~Piekarewicz, {Constraining URCA cooling of neutron stars from the neutron radius of Pb-208}, Phys. Rev. C 66 (2002) 055803.
\newblock \href {http://arxiv.org/abs/nucl-th/0207067} {\path{arXiv:nucl-th/0207067}}, \href {https://doi.org/10.1103/PhysRevC.66.055803} {\path{doi:10.1103/PhysRevC.66.055803}}.

\bibitem{Dexheimer:2018dhb}
V.~Dexheimer, R.~de~Oliveira~Gomes, S.~Schramm, H.~Pais, {What do we learn about vector interactions from GW170817?}, J. Phys. G 46~(3) (2019) 034002.
\newblock \href {http://arxiv.org/abs/1810.06109} {\path{arXiv:1810.06109}}, \href {https://doi.org/10.1088/1361-6471/ab01f0} {\path{doi:10.1088/1361-6471/ab01f0}}.

\bibitem{Dexheimer:2020rlp}
V.~Dexheimer, R.~O. Gomes, T.~Kl\"ahn, S.~Han, M.~Salinas, {GW190814 as a massive rapidly rotating neutron star with exotic degrees of freedom}, Phys. Rev. C 103~(2) (2021) 025808.
\newblock \href {http://arxiv.org/abs/2007.08493} {\path{arXiv:2007.08493}}, \href {https://doi.org/10.1103/PhysRevC.103.025808} {\path{doi:10.1103/PhysRevC.103.025808}}.

\bibitem{Fattoyev:2010rx}
F.~J. Fattoyev, J.~Piekarewicz, {Relativistic models of the neutron-star matter equation of state}, Phys. Rev. C 82 (2010) 025805.
\newblock \href {http://arxiv.org/abs/1003.1298} {\path{arXiv:1003.1298}}, \href {https://doi.org/10.1103/PhysRevC.82.025805} {\path{doi:10.1103/PhysRevC.82.025805}}.

\bibitem{Fukushima:2003fw}
K.~Fukushima, {Chiral effective model with the Polyakov loop}, Phys. Lett. B 591 (2004) 277--284.
\newblock \href {http://arxiv.org/abs/hep-ph/0310121} {\path{arXiv:hep-ph/0310121}}, \href {https://doi.org/10.1016/j.physletb.2004.04.027} {\path{doi:10.1016/j.physletb.2004.04.027}}.

\bibitem{Clevinger:2022xzl}
A.~Clevinger, J.~Corkish, K.~Aryal, V.~Dexheimer, {Hybrid equations of state for neutron stars with hyperons and deltas}, Eur. Phys. J. A 58~(5) (2022) 96.
\newblock \href {http://arxiv.org/abs/2205.00559} {\path{arXiv:2205.00559}}, \href {https://doi.org/10.1140/epja/s10050-022-00745-3} {\path{doi:10.1140/epja/s10050-022-00745-3}}.

\bibitem{Aryal:2020ocm}
K.~Aryal, C.~Constantinou, R.~L.~S. Farias, V.~Dexheimer, {High-Energy Phase Diagrams with Charge and Isospin Axes under Heavy-Ion Collision and Stellar Conditions}, Phys. Rev. D 102~(7) (2020) 076016.
\newblock \href {http://arxiv.org/abs/2004.03039} {\path{arXiv:2004.03039}}, \href {https://doi.org/10.1103/PhysRevD.102.076016} {\path{doi:10.1103/PhysRevD.102.076016}}.

\bibitem{Yao:2023yda}
N.~Yao, A.~Sorensen, V.~Dexheimer, J.~Noronha-Hostler, {Structure in the speed of sound: from neutron stars to heavy-ion collisions} (11 2023).
\newblock \href {http://arxiv.org/abs/2311.18819} {\path{arXiv:2311.18819}}.

\bibitem{Roark:2018uls}
J.~Roark, V.~Dexheimer, {Deconfinement phase transition in proto-neutron-star matter}, Phys. Rev. C 98~(5) (2018) 055805.
\newblock \href {http://arxiv.org/abs/1803.02411} {\path{arXiv:1803.02411}}, \href {https://doi.org/10.1103/PhysRevC.98.055805} {\path{doi:10.1103/PhysRevC.98.055805}}.

\bibitem{Dexheimer:2020okt}
V.~Dexheimer, K.~Aryal, .~M. Wolf, .~C. Constantinou, R.~L.~S. Farias, {Deconfinement Phase Transition under Chemical Equilibrium}, Astron. Nachr. 342~(1-2) (2021) 347--351.
\newblock \href {http://arxiv.org/abs/2011.11686} {\path{arXiv:2011.11686}}, \href {https://doi.org/10.1002/asna.202113932} {\path{doi:10.1002/asna.202113932}}.

\bibitem{Aryal:2021ojz}
K.~Aryal, C.~Constantinou, R.~L.~S. Farias, V.~Dexheimer, {The Effect of Charge, Isospin, and Strangeness in the QCD Phase Diagram Critical End Point}, Universe 7~(11) (2021) 454.
\newblock \href {http://arxiv.org/abs/2109.14787} {\path{arXiv:2109.14787}}, \href {https://doi.org/10.3390/universe7110454} {\path{doi:10.3390/universe7110454}}.

\end{thebibliography}
\end{document}